\documentclass[conference,psfig,10pt,a4paper]{IEEEtran}
\IEEEoverridecommandlockouts
\usepackage[left=1.60cm,right=1.60cm,top=2.20cm,bottom=3.20cm]{geometry}
\usepackage{cite}
\usepackage{amsmath,amssymb,amsfonts}
\usepackage{algorithmic}
\usepackage{graphicx}
\usepackage{textcomp}
\usepackage{xcolor}
\def\BibTeX{{\rm B\kern-.05em{\sc i\kern-.025em b}\kern-.08em
    T\kern-.1667em\lower.7ex\hbox{E}\kern-.125emX}}

\begin{document}

\title{Relay-aided Slotted Aloha for Optical Wireless Communications}

\author{\IEEEauthorblockN{Milica Petkovic, Dejan Vukobratovi\'{c}}
\IEEEauthorblockA{University of Novi Sad \\
Faculty of Technical Sciences \\
21000 Novi Sad, Serbia\\
Emails: $\lbrace$milica.petkovic, dejanv$\rbrace$@uns.ac.rs}
\and
\IEEEauthorblockN{Andrea Munari, Federico Clazzer}
\IEEEauthorblockA{Institute of Communications and Navigation of \\
the German Aerospace Center (DLR) \\
82234 Wessling, Germany\\
Emails: $\lbrace$andrea.munari, federico.clazzer$\rbrace$@dlr.de}
}

\maketitle

\begin{abstract}
We consider a relay-aided Slotted ALOHA solution for uplink random access for an Optical Wireless Communications (OWC)-based Internet of Things (IoT). The first phase of uplink, the one between IoT devices and the relays, is realized using indoor OWC, while the second phase, between the relays and a base station, represents the long-range RF transmission$~$based on low-power wide area network such as LoRaWAN and occurs outdoors. The throughput performance dependence on the OWC and RF channel conditions is observed. The behavior of the performance gain due to adding relays is highlighted and investigated under different channel and traffic conditions.
\end{abstract}


\section{Introduction}
With constant increase of the number of user devices, the upcoming generations of wireless technologies are faced with demanding requirements in order to satisfy their connectivity for Internet of Things (IoT) applications. One of the main challenges of 5G and beyond-5G systems is to provide reliable communications with high energy- and spectral-efficiency \cite{mag}. Due to short packet transmission and unpredictability of device activity, the conventional approach to address this problem is to use random access (RA) protocols. In RA-based networks, the user nodes share the communication medium in an uncoordinated manner, avoiding the costs of  resource allocation  \cite{mag, from5to6}. Among different RA approaches, simple ALOHA  based RA schemes \cite{SA,Roberts,Abramson} have been adopted in commercial systems \cite{lora,sigfox,3gpp}. Furthermore, the slotted ALOHA (SA) approach with spatial diversity was studied in \cite{zorzi, SD1, SD4, SD5}, assuming that users generate traffic which can be detected by  multiple receivers. This multi-receiver SA  setup is complemented in \cite{SD2, SD3, balkancom}, by considering the two-tier topology  where multiple receivers act as relays and forward recovered packet to a common sink. More precisely, time-division multiple access is employed for the relays-to-sink links in \cite{SD2, SD3}, while \cite{balkancom} considers the SA policy for the same link. 
%The independent erasure channel (on-off fading) model is assumed for both users-to-relays and relays-to-sink links in   \cite{SD2, SD3, balkancom}.

Although most of the considered  IoT  applications  are related  to conventional radio frequency (RF) technologies, recently there has been particular  attention directed to  optical wireless communication (OWC) as a 5G wireless  emerging  technology. Indoor OWC systems operating  in  visible spectrum,  called visible light communications (VLC), represent  
an efficient alternative to the RF systems, since they provide very high-speed, green and secure   transmission \cite{LOS,model,OWC_MATLAB,bookvlc}.
The VLC-based indoor IoT systems have been analyzed in \cite{bookvlc,vlc1,vlc2,vlc3,vlc4}. In order to ensure throughput-efficient link in optical IoT networks, different multiple
access approaches have been analyzed  \cite{MA1,MA2,MA3,MA4}. Recently, the SA policy is adopted for OWC-based IoT  systems in \cite{SA1,SA2}, where the uplink VLC system is analyzed considering multi-packet reception (MPR) and successive interference cancellation.

Motivated by aforementioned, in this paper the OWC-based two-tier SA multiple-relay system is analyzed. We consider a similar scenario as in \cite{balkancom}, i.e., the uplink represents the data  transmission from users to the relays, which further forward the packets to the common sink. 
%The SA policy is employed within this two hop system considering simple on-off fading channels for both uplink and downlink. 
Differently from \cite{balkancom}, we propose the scenario where the first phase of uplink represents the SA-based OWC signal  transmission from the IoT user devices to finite number of decode-and-forward (DF) relays in indoor environment. In the second  phase of the uplink, the recovered data packets are further transmitted from the relays to the base station following a SA method. The second phase is performed in outdoor environment and across possibly larger distances, which makes Long Range Wide Area Network (LoRaWAN) an ideal candidate, providing license-free RF-based long-range communication combined with low energy consumption \cite{lora,lora1,lora2,lora3}. Unlike \cite{balkancom} which considers simple on-off fading channels for both uplink and downlink, in this paper, the erasure events will happen when the power contribution of the received packets is lower than a previously determined power threshold. This model is then integrated into an exact expression of the end-to-end throughput and presented assuming no MPR and no capture effect. The resulting expression is used to observe the trade-offs between the system throughput performance and the system parameters.

The rest of the paper is organized as follows. Section II presents the system model, while the end-to-end  throughput  performance analysis are provided in Section III. Numerical results and  discussions are given in Section IV. Section V concludes the paper.

\section{System Model}

\begin{figure}[b]
\centerline{\includegraphics[width=3.5in]{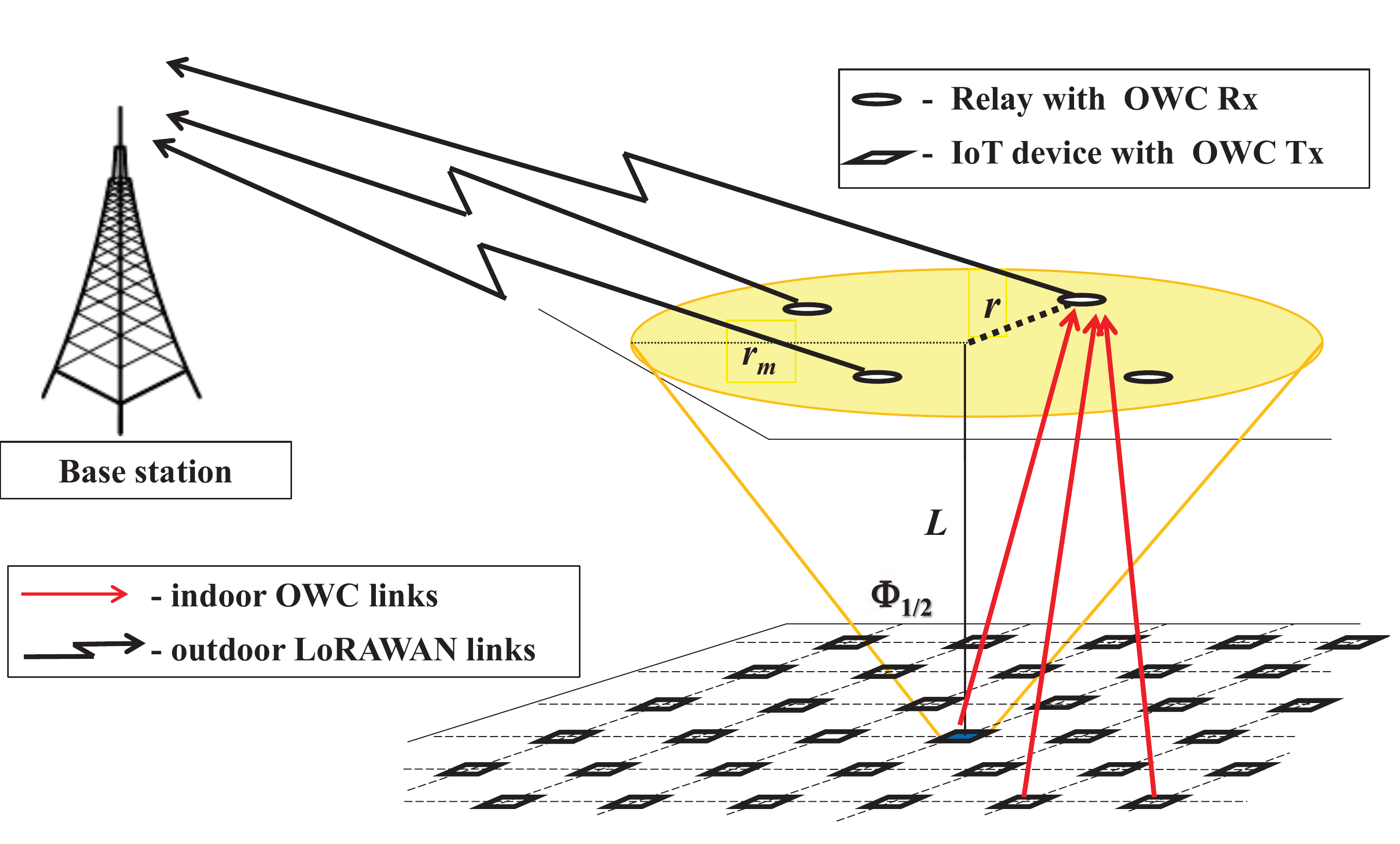}}
\caption{Considered scenario.}
\label{Fig1}
\end{figure}

The two-tier topology of the considered system  is presented in Fig.~\ref{Fig1}. Transmission is divided into two phases, where the first one is related to the uplink indoor OWC scenario, while the second phase represents the LoRAWAN outdoor transmission.
The SA medium access protocol is adopted for both indoor and outdoor transmissions \cite{Roberts}. 

%\newcounter{MYtempeqncnt}
%\begin{figure*}[!t]
%\normalsize
%%\setcounter{MYtempeqncnt}{\value{equation}}
%%\setcounter{equation}{8}
%\centering
%\includegraphics[width=5.5in,height=7cm]{Fig1.pdf}
%\caption{System model.}
%\label{Fig_1}
%
%%\setcounter{equation}{\value{MYtempeqncnt}}
%\vspace*{4pt}
%\end{figure*}

\subsection{The indoor OWC phase}

In the first phase, the IoT devices perform data packet transmission via OWC transmitters, usually LED lamps operating in visible or IR spectrum. Data packets are sent in an uncoordinated fashion to $K$ DF-based relays placed on the ceiling of the indoor space.  
The intensity modulation with on-off keying is utilized in order to satisfy non-negative constraint.
All relays contain the OWC receivers (photo detectors) where direct detection and optical-to-electrical signal conversion is performed. 
% The assumption that the LEDs footprints cover all of implemented relay is adopted. 

% As it can be observed in Fig.~\ref{model}, distance between the LED  transmitter and photodetector receiver planes is denoted by $L$, i.e., IoT devices are positioned in the plane at height $ L $  from the ceiling. Furthermore, the user devices are randomly placed at the floor plane at the distance $ L $ from the ceiling. 

Even though optical wireless links include both LoS and diffuse components, 
the reflected signals energy can be  neglected since it is proved to be significantly lower than the energy of the LoS component  \cite{LOS}.
Following the Lambertian law for modeling the optical LoS link, the intensity of the optical signal between the  user and the  relay is defined as \cite{OWC_MATLAB,bookvlc}
\begin{equation}
 I =\!\! \left\{ {\begin{array}{*{20}{c}} \!\!\!\!\!
\frac{\mathcal A\left( m + 1\right)\mathcal R T_s g\left( \psi \right)}{2\pi d^2}\cos ^m\left( \theta \right)\cos \left( \psi \right), 0 \leq \psi \leq \Psi\\
{0,}  \quad \quad \quad \quad \quad \quad \quad \quad \quad \quad\quad\quad {\rm otherwise} \\
\end{array}} \right.\!\!\!\!,
\label{I_ij}
\end{equation}
where the parameter $\Psi$ denotes the field of view (FOV) of the receiver. We assume that the FOV of any of the OWC receivers is sufficiently large to detect the signal from any of the IoT devices, conditioned that the received power is sufficiently high. Parameters  $ d$, $\theta$ and $\psi$ are the Euclidean distance between transmitter and receiver, the irradiance angle and the incidence angle, respectively.  The  physical surface area of the photodetector is denoted by $ \mathcal A $, $\mathcal R$ is the responsivity, and $T_s$ defines the gain of the optical filter. The optical concentrator is determined as $g\left( \psi \right) =\zeta^2 /\sin^2\left( \Psi \right)$, for $0\leq\psi \leq \Psi$, where $\zeta$ represents the refractive index of lens at a photodetector. The OWC transmission  follows a generalized Lambertian radiation pattern with the order $m$ as \cite{OWC_MATLAB}
\begin{equation}
m = - \frac{\ln 2}{\ln \left( \cos \Phi_{1/2} \right)},
\label{m}
\end{equation}
where  $\Phi_{1/2}$ represents the semi-angle at the half illuminance of LED, and defines the width of the optical beam. 
From Fig.~\ref{Fig1} can be concluded that the  semi-angle at the half illuminance of LED is related to the maximum radius of a LED lighting footprint, $r_{m}$, as   $r_{m} =  L\tan \left(\Phi_{1/2} \right) $, where $L$ represents the distance  between horizontal plane where users are located on the ceiling.
With the assumption that the surface of OWC transmitters is parallel to the ceiling plane and there is no orientation of the OWC receivers, then $\theta=\psi $,   $d=~\sqrt {r^2 + L^2} $,    $\cos \left( \theta \right) \!=\! \frac{L}{ \sqrt {r^2 + L^2} }$, where $r$ represents the distance between relay and projection of the user position on the ceiling plane. The expression in (\ref{I_ij}) for the optical signal intensity can be  rewritten as
\begin{equation}
 I =\!\! \left\{ {\begin{array}{*{20}{c}} \!\!\!\!\!
\frac{\mathcal X  }{\left( r^2 + L^2 \right)^{\frac{m + 3}{2}}}, \quad \quad\quad0 \leq \psi \leq \Psi\\
{0,}  \quad  \quad  \quad \quad\quad\quad {\rm otherwise} \\
\end{array}} \right.\!\!\!\!,
\label{I_ij_1}
\end{equation}
where  $\mathcal X = \frac{\mathcal A\left( m + 1 \right)\mathcal R}{2\pi}T_sg\left( \psi \right)L^{m + 1}$.

The user devices are randomly placed at the floor plane, while the $K$ relays are at fixed positions. 
If the positions of IoT  devices are  on the same plane and modeled by a uniform distribution, the probability density function (PDF) of the  radial distance $r$ of a randomly placed user from a fixed receiver is \cite{model}
\begin{equation}
f_{r}\left( r \right) = \frac{2r}{r_{m}^2},\quad 0\leq r \leq r_{m}.
\label{pdf_rn}
\end{equation}
 After utilization of the  technique for transformation of random variables, based on (\ref{I_ij_1}) and (\ref{pdf_rn}), the PDF of the optical signal intensity is expressed as \cite{model}
\begin{equation}
f_{ I}\left(  I \right) = \frac{2 \mathcal X ^{\frac{2}{m + 3}}}{r_{m}^2\left( m + 3 \right)}  I^{ - \frac{m + 5}{m + 3}},\quad  I_{\min }\leq  I \leq  I_{\max },
\label{pdf_h}
\end{equation}
where $ I_{\min} = \frac{\mathcal X }{{\left( r_{m}^2 + L^2 \right)}^{\frac{m + 3}{2}}}$ and  $I_{\max} = \frac{\mathcal X}{L^{m + 3}}$.
The instantaneous SNR of the VLC link can be defined as \cite{OWC_MATLAB, model}
\begin{equation}
\gamma_{\rm vlc} = \frac{P_t^2 I^2 \eta ^2}{N_0 B}, 
\label{snrVLC}
\end{equation}
where $P_t$ denotes the average transmitted optical power of a LED lamp, $\eta$ represents  optical-to-electrical conversion
efficiency, $N_0$ is noise spectral density and $B$ is the system  bandwidth.  

Based on the instantaneous SNR in (\ref{snrVLC}) and the PDF of optical signal intensity in (\ref{pdf_h}),  the PDF of the instantaneous SNR  is derived as 
\begin{equation}
f_{\gamma_{\rm vlc}}\left( \gamma  \right) = \frac{\left( \mu _{\rm vlc}\mathcal X ^2 \right)^{  \frac{1}{m + 3}}}{r_{m}^2\left( m + 3 \right)}\gamma^{ - \frac{m + 4}{m + 3}},\quad \gamma_{\min }\leq \gamma \leq \gamma_{\max },
\label{pdfVLC}
\end{equation}
where $\gamma_{\min} = \frac{\mu_{\rm vlc}  \mathcal X^2}{{\left( r_{m}^2 + L^2 \right)}^{m + 3}}$ and  $\gamma_{\max} = \frac{\mu_{\rm vlc} \mathcal X^2}{L^{2 \left(m + 3\right)}}$, and
\begin{equation}
\mu_{\rm vlc}=\frac{P_t^2\eta ^2}{N_0 B}.
\label{avSNR_VLC}
\end{equation}
Furthermore, after performing integration, the cumulative distribution function (CDF) of the instantaneous SNR is 
\begin{equation}
F_{\gamma_{\rm vlc}}\!\!\left( \gamma  \right)\!\! =\!\! \left\{ {\begin{array}{*{20}{c}}
\!\!\!{1\! +\! \frac{L^2}{r_{m}^2}\! - \!\frac{1}{r_{m}^2}{{\left( {\frac{ \mu _{\rm vlc}\mathcal X ^2  }{\gamma}} \right)}^{\frac{1}{{m + 3}}}},}  \gamma_{\min}\!\!\leq \!\!\gamma \leq\gamma_{\max}\\
{1,}  \quad \quad \quad \quad \quad \quad \quad \quad \quad \quad \gamma > \gamma_{\max}\\
\end{array}} \right.\!\!\!\!.
\label{cdfVLC}
\end{equation}

\subsection{The  outdoor LoRaWAN phase}

The second phase of two-tier random access scenario   in Fig.~\ref{Fig1} represents the RF uplink channel between relays and the base station. Upon correctly receiving the information data, each DF relay re-encodes and forwards data packets\footnote{SA approach and  data recovery at relays and base station will be discussed in the next section.}. It is assumed that no buffering is done at the relays, thus the packets are either sent to the base station in the subsequent time slot, or they are discarded, as we discuss later. Since the signal transmission is performed in the outdoor scenario, the SA-based LoRaWAN technology is adopted in this phase as a low power and licence-free transmission. 
The instantaneous SNR over RF link is defined as  
\begin{equation}
\gamma_{\rm rf} = \frac{h^2 P_s}{\sigma _R^2}, 
\label{snrRF}
\end{equation}
where $h$  is the signal fading amplitude over RF link, $P_s$ represents an average transmitted  power and $\sigma _R^2$ is the the additive white Gaussian noise variance. The average SNR is defined as
\begin{equation}
\mu_{\rm rf} = \mathbb E \left[ \gamma_{\rm rf} \right] = \mathbb E \left[ \frac{h^2 P_s}{\sigma _R^2} \right] . 
\label{AVsnrRF}
\end{equation}
We assume that each relay-to-base station RF channel follows independent and identical distributed (i.i.d.) Nakagami-\textit{m} distribution, which is a general model suitable for both LoS and non-LoS transmissions \cite{mmW3}.
The PDF and the CDF of the instantaneous SNR of each link are given respectively as \cite{Alouni}
\begin{equation} \label{pdfRF}
f_{\gamma_{\rm rf}}\!\left( \gamma \right) \!= \frac{{{m_1}^{m_1}}{\gamma^{m_1 - 1}}}{{{\mu_{\rm rf} ^{m_1}}\Gamma \left( m_1 \right)}}e^{- \frac{{m_1\gamma}}{\mu_{\rm rf} }},
\end{equation}
\begin{equation} \label{cdfRF}
F_{\gamma_{\rm rf}}\!\left( \gamma\right)\! =\!1\! - \!\frac{\Gamma \left( {m_1,\frac{{m_1\gamma}}{\mu_{\rm rf} }} \right)}{{\Gamma \left( m_1 \right)}},
\end{equation}
where $m_1$ is the Nakagami-\textit{m} fading parameter, and $\Gamma \left( \cdot, \cdot \right)$ denotes the Incomplete Gamma function defined in \cite[
(8.350.2)]{grad}.

\section{Throughput Performance Analysis}

%\newcounter{MYtempeqncnt}
\begin{table*}[t]

\normalsize
\centering
\label{tab:label}
\caption{\cite{model,vlc1a, Haas}}

\begin{tabular}{||c|c|c||}
\hline \hline
{name} & {symbol} & {value}\\
    
\hline\hline
Field of view (FOV) of the
receiver	& $ \Psi $ &	$ 90^{\circ} $   \\

Photodetector
surface area  &  $\mathcal{A}$ &	$1 $ cm${^2}$\\

Responsivity & $\mathcal{R}$ &	$ 0.4 $ A/W  \\

Optical filter gain  & $T_s $ &  $	$  1 \\

Refractive index of lens at a photodetector	& $ \zeta $ &$ 1.5$  \\

Optical-to-electrical conversion efficiency & $ \eta $ &	$ 0.8 $  \\

Noise power
spectral density& $ N_{\rm 0} $ &	$ 10^{-21} $ W/Hz \\

System bandwidth & $B $ &$ 20 $ MHz \\

\hline\hline
\end{tabular}

% The spacer can be tweaked to stop underfull vboxes.
\vspace*{6pt}

%\hrulefill

\end{table*}

As mentioned, the SA protocol is adopted as a medium access policy, with the assumption that the users are slot-synchronized. Transmission can start only at the beginning of time slots and one packet occupies exactly one slot. We name the  average number of packets sent per slot  as the channel load $G$ [pk/slot]. The number of users generating the data over a slot can be modeled by  the binomial distribution, which  can be well approximated with the Poisson distribution.
Hence, we assume that the number of users generating the data over a slot are modeled  a Poisson-distributed random variable $U$ with intensity $G$ as 
\begin{equation}
{\rm Pr} \lbrace U =u \rbrace = \frac{G^u e^{-G}}{u!}.
\label{Poisson}
\end{equation}
Following a SA policy with assumption that there is no MPR capabilities and no capture effect  at the relay, the data will be retrieved only if a single packet reaches the relay during a slot \cite{balkancom}. In other words, collisions are considered to be destructive, and the relay will recover the data only if one of the $U = u$ transmitted packets reaches the receiver.

In the user-to-relay transmission link, we adopt the  model which assumes that the packet at the relay will be erased if the instantaneous SNR over VLC link, $\gamma_{\rm vlc}$, defined in (\ref{snrVLC}), is lower than previously determined threshold $\gamma^{(1)}_{\rm th}$. 
 In other words, the erasure event will happen if the total power contribution of received packet at the relay is lower than the power threshold which is in relation  with  $\gamma^{(1)}_{\rm th}$. Thus,  a packet at the relay is either in deep fade, i.e., erased, with probability $\epsilon_{\rm vlc}$, or it arrives unfaded with probability $1 - \epsilon_{\rm vlc}$. Since the user-to-relay link represents the OWC channel (being characterized by the optical signal intensity defined in (\ref{I_ij})), and the OWC users are  randomly placed on the floor in the room, the probability $\epsilon_{\rm vlc}$ is equal to the CDF defined in (\ref{cdfVLC}), i.e., $\epsilon_{\rm vlc}= F_{\gamma_{\rm vlc}} \left( \gamma^{(1)}_{\rm th} \right) $.  Hence, the  successful reception of data conditioned on $U - u$ transmission occurs with probability 
\begin{equation}
{\rm p}_u : = u (1-\epsilon_{\rm vlc})\epsilon_{\rm vlc}^{u-1}.
\label{pu}
\end{equation}
The average throughput experienced at each of the $K$ relays, in terms of average decoded packets per slot, can be determined after removing the conditioning as
\begin{equation}
S_{\rm up}= \sum_{u=1}^{\infty} \frac{G^u e^{-G}}{u!}  {\rm p}_u = G(1 - \epsilon_{\rm vlc}) e ^{-G(1 - \epsilon_{\rm vlc})},
\label{Tup}
\end{equation}
which corresponds to the throughput of a SA link with erasures.

In next uplink LoRaWAN phase, $K$ relays contend to send the recovered data to the base station through a SA policy. We adopt the scheme proposed in \cite{balkancom} where, after decoding the successfully recovered packets, the relays independently determine if the packet will be forwarded. The probability if data will be transmitted further to the base station in the subsequent slot is denoted with $\delta$, while $1 - \delta$ holds for the probability that a data will be discarded. Similarly as in the OWC uplink, we assume that there are no MPR and no capture effect  at the base station, thus the data will be correctly decoded if only a single packet reaches the base station during a slot. As it was mentioned in Section II, the  RF channels  between relays and base station  are characterized as  i.i.d. Nakagami-\textit{m} fading links. The data packet at the base station will be erased with probability $\epsilon_{\rm rf}$, which is in relation with the CDF defined in (\ref{cdfRF}) as $\epsilon_{\rm rf} = F_{\gamma_{\rm rf}} \left( \gamma^{(2)}_{\rm th} \right)$, where 
$\gamma^{(2)}_{\rm th} $ is previously determined   threshold.

In order to derive the end-to-end throughput of the system under investigation, we follow the approach presented in \cite{balkancom}. We first we define the probability 
\begin{equation}
{\rm q}_u : = {\rm p}_u \delta (1- \epsilon_{\rm rf}),
\label{qu}
\end{equation}
which represents the overall probability that relay successfully decodes one of the packets sent by users, forwards it with probability $\delta$,   and the corresponding data is received unfaded at the base station. Information will be recovered  only
if a single packet is received at the base station  unfaded, while the rest of them are erased. This is defined by the binomial probability  conditioned on $U=u$ as
\begin{equation}
{\rm z}_u := K {\rm q}_u  (1 - {\rm q}_u)^{K - 1}.
\label{zu}
\end{equation}
Finally, the end-to-end throughput for the system under investigation can then be determined by removing conditioning on ${\rm z}_u$ as
\begin{equation}
S= \sum_{u=1}^{\infty} \frac{G^u e^{-G}}{u!}  {\rm z}_u =\sum_{u=1}^{\infty} \frac{G^u e^{-G}}{u!} K {\rm q}_u  (1 - {\rm q}_u)^{K - 1}.
\label{T}
\end{equation}

Since the end-to-end throughput expression in (\ref{T}) is presented in terms of infinite series, following the derivation presented in \cite[Appendix A]{balkancom}, the closed-form expression of  end-to-end throughput is derived as \cite{balkancom} 
\begin{equation}
\begin{split}
S &= \sum_{i=0}^{K-1} (-1)^i K {K-1 \choose i}e^{-G} \\
& \times \left(\frac{ \delta(1 - \epsilon_{\rm vlc}) (1 - \epsilon_{\rm rf}) }{\epsilon_{\rm vlc}} \right) ^ {i + 1} H_{i+1} \left( G \epsilon_{\rm vlc}^{i+1} \right),
\label{T3}
\end{split}
\end{equation}
where  the ancillary function  is defined as
\begin{equation}
\begin{split}
H_m (x) = \left\{ {\begin{array}{*{20}{c}} \quad \quad
~~e^x, \quad \quad\quad \quad\quad \quad m=0 \\
x \sum_{l=0}^{m-1} {m-1 \choose l} H_l (x), \quad     m\geq 1 \\
\end{array}}. \right. 
\label{H}
\end{split}
\end{equation}

Note that, although the  structure of our and the system model proposed in \cite{balkancom} is the same, the results are different since we observe more general channel conditions compared to \cite{balkancom}. Also, unlike \cite{balkancom} which is inspired by satellite communications, this paper considers different technologies applied in the two phases of the uplink transmission. 

\section{Numerical Results}

In this section,  numerical results are presented which are obtained by using derived
analytical expressions for the end-to-end throughput.
Table I represents the  values of the parameters  assumed in this section \cite{model,vlc1a, Haas}. 
It is assumed that the SNR thresholds are the same for both indoor and outdoor  uplinks, i.e., $\gamma^{(1)}_{\rm th}=\gamma^{(2)}_{\rm th}=\gamma_{\rm th}$. The probability that the data at the relay is directly forwarded to the base station is equal to $\delta=1$.

\begin{figure}[!t]
\centerline{\includegraphics[width=3.5in]{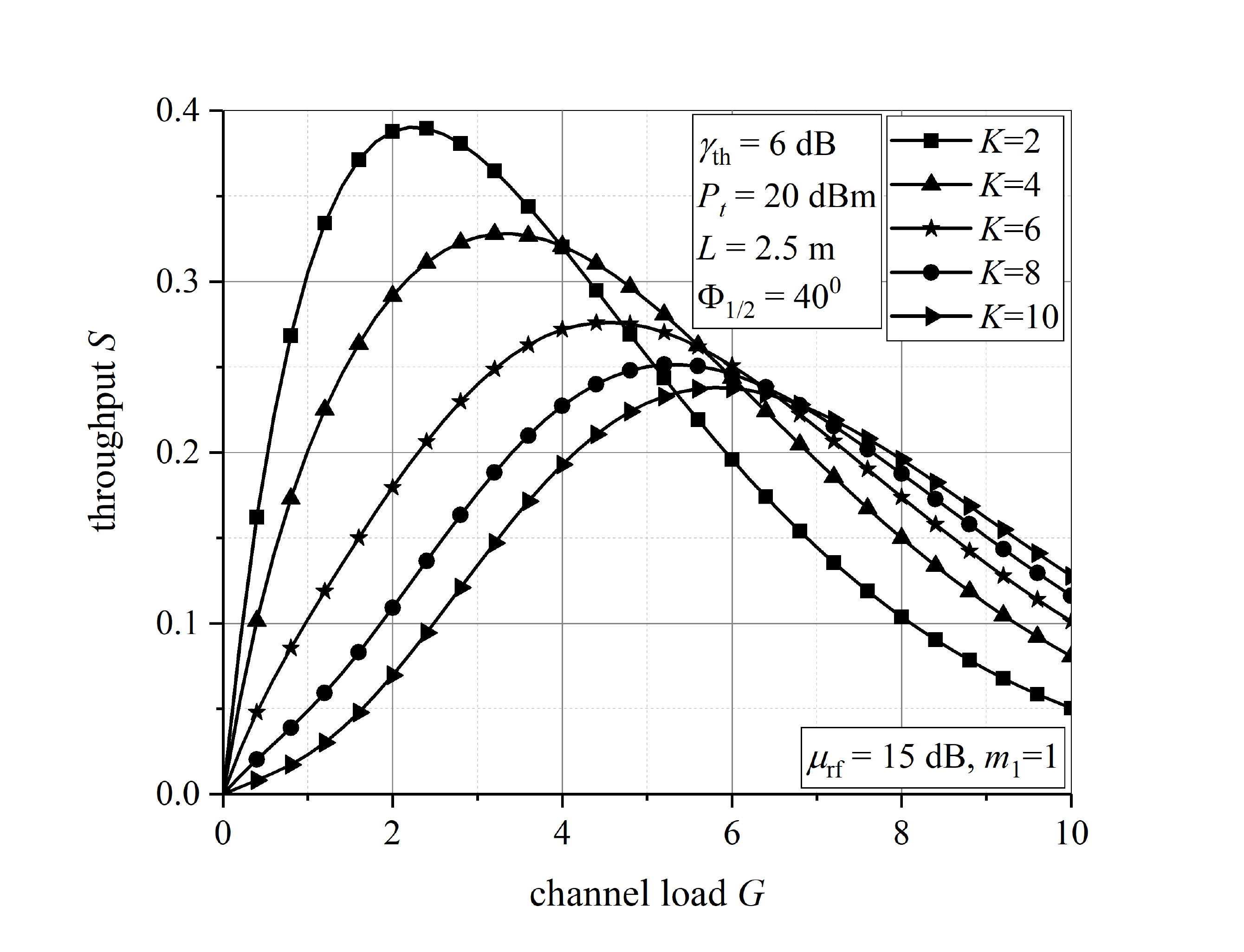}}
\caption{End-to-end throughput vs. channel load for different number of relays.}
\label{Fig2}
\end{figure}
\begin{figure}[t]
\centerline{\includegraphics[width=3.5in]{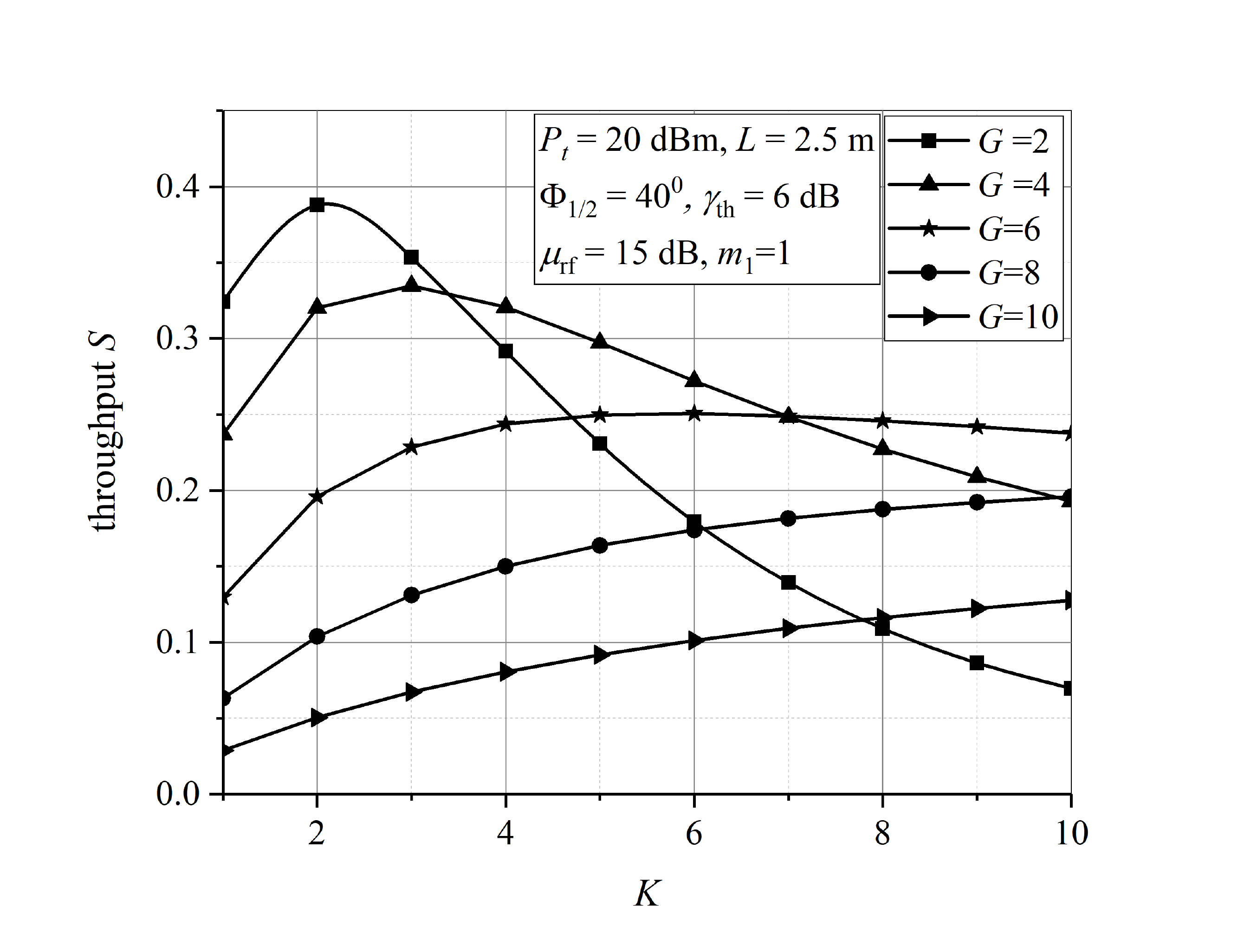}}
\caption{End-to-end throughput vs.  number of relays.}
\label{Fig3}
\end{figure}

Fig.~\ref{Fig2} depicts the end-to-end throughput dependence on the channel load, when different number of employed relays are  considered.  For lower channel load $G$, greater number of relays leads to performance deterioration. This happens since we have fewer users, and employing more relays  results in the situation when more packets will have  power contribution  higher than power threshold (resulting in collisions).
Furthermore, the maximal value of the system throughput is observed   for optimal value of channel load,  $G_{\rm opt}$. This optimal value   is dependent on the number of active relays. When $K$ is higher, the optimal value $G_{\rm opt}$ is also higher, but the value of the maximal  throughput is lower. 
After achieving its maximal value, the end-to-end throughput will be reduced with further increasing of $G$. In this areas of intensive load, more users contend for the same number of relays, thus the probability that the collision will happen is increased.

\begin{figure}[t]
\centerline{\includegraphics[width=3.5in]{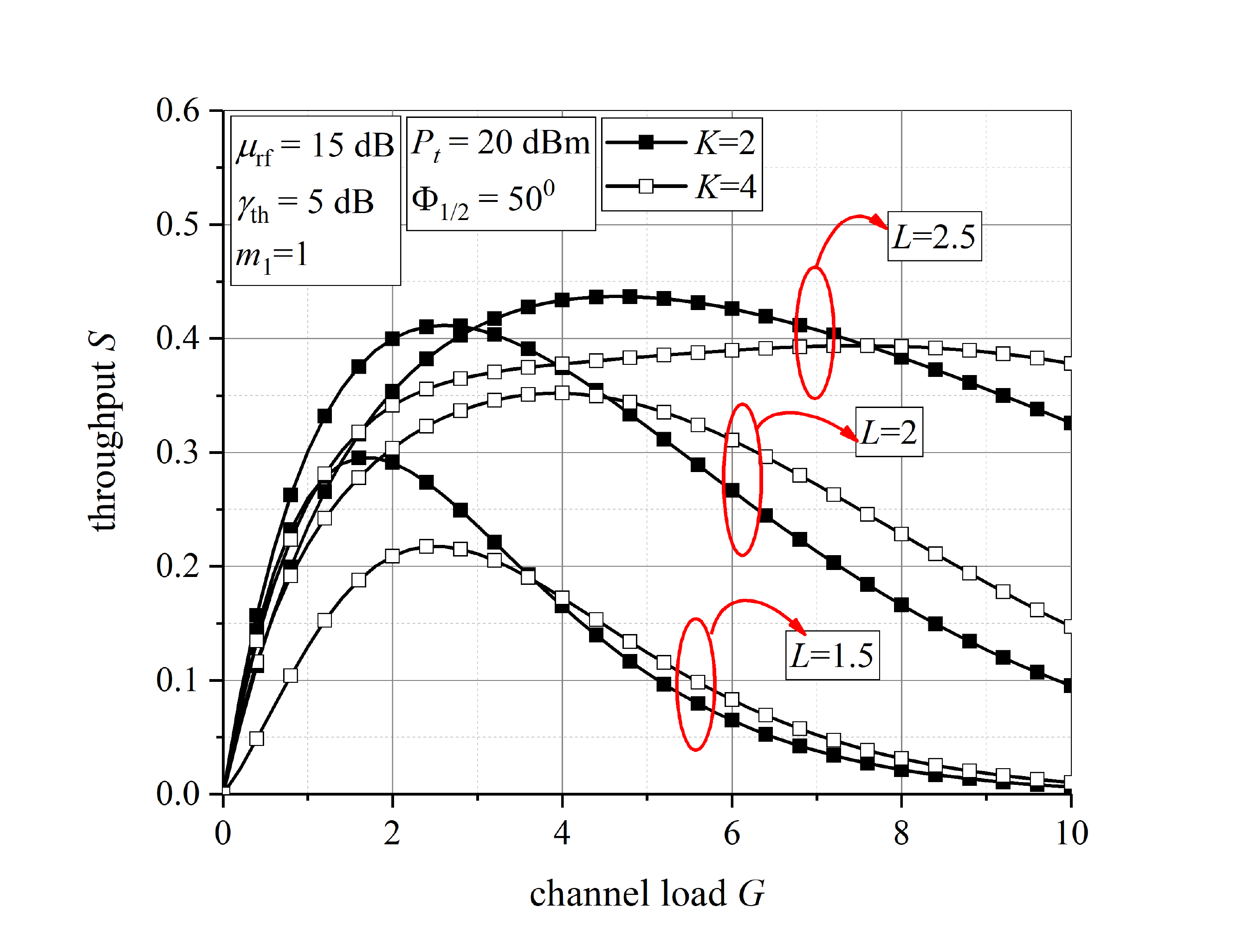}}
\caption{End-to-end throughput vs. channel load for different values of room height.}
\label{Fig4}
\end{figure}

Similarly,  Fig.~\ref{Fig3} presents how the increase in the number of relays affects the system throughput for different load conditions. For lower channel load, adding the relays  will cause throughput performance improvement until some point. After this optimal number of relays, appending more relays will cause system performance impairment since most of them will overpower
 the  threshold and there will  be a lot of unfaded  packets at the relays. The collisions will occur resulting in lower throughput. 
For higher $G$, more users access the shared medium, thus greater number of relays  reflects in the performance improving. 

The end-to-end throughput dependence on the channel load for different room heights is presented in Fig.~\ref{Fig4}. The systems employing $K=2$ and $K=4$ are considered. The maximal values of throughput are also observed for different optimal values of channel load. With greater distance between planes where IoT users and relays are placed, the value of $G_{\rm opt}$ is higher. It can be concluded that the system throughput is greater with greater $L$. The OWC channel conditions are more convenient for signal transmission when $L$ is smaller, since optical signal intensity (and received power) is stronger when optical signal propagates through shorter distances. Still, the system throughput for considered scenario is reduced with  lower $L$ since a lot of packets will be received with enough power to overcome threshold, thus they will not be erased and collisions will occur.

\begin{figure}[t!]
\centerline{\includegraphics[width=3.5in]{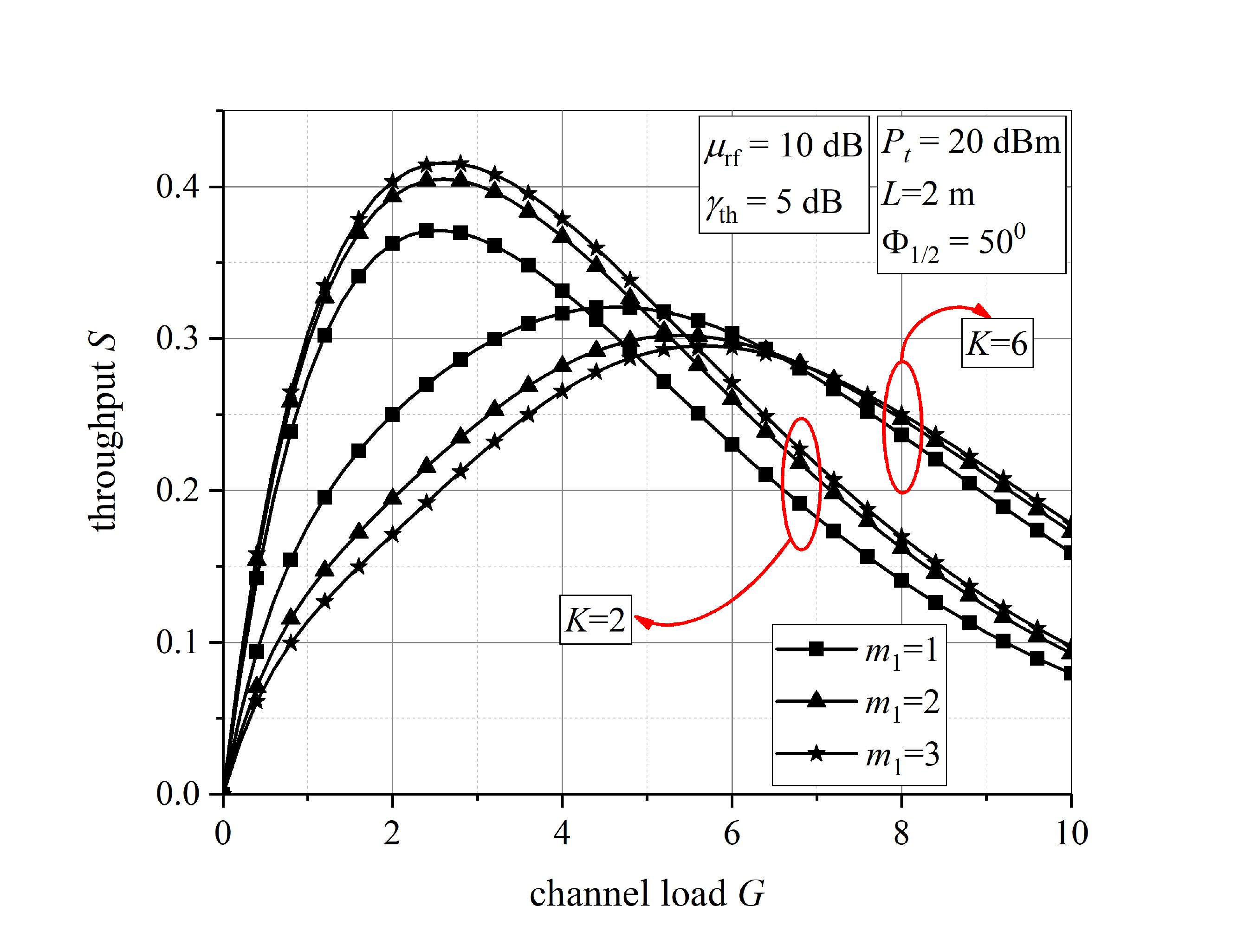}}
\caption{End-to-end throughput vs. channel load for different values of Nakagami-\textit{m} fading parameter.}
\label{Fig5}
\end{figure}

Fig.~\ref{Fig5} shows the end-to-end throughput versus  channel load, considering different fading severity for RF channels. When system with $K=2$ relays is considered, it is noticed that lower $m_1$, meaning that the fading is stronger, results in worse system performance. Since there are only two relays, at most  two packets can be forwarded to the base station over a slot. The probability to have collided packets is small, since collision will occur when both relays recover a single packet over previous slot. In that case, fading has important role in determining the system throughput. With worsening RF channel conditions (lower $m_1$), throughput will be reduced since power of received power will be lower leading to the higher probability that the packet is erased. 
On the other hand, when  $K=4$, the inverse effect is noticed for lightly loaded conditions. Since now the most four packets can be transmitted to the base station from relays, stronger fading can reduce the power contributions of sent packets, resulting in the case that some of them can be erased, which will improve  throughput. After achieving the maximal value, system throughput will be decreased. In this area, better RF channel conditions reflects in better  throughput.

\section{Conclusion}
In this work, we have presented a multireceiver SA based two-tier system. The first part of transmission is the indoor OWC uplink, where the IoT  devices communicate with multiple relays by SA approach. The second phase refers to the LoRaWAN based on SA transmission  in outdoor environment. The end-to-end throughput is determined based on the packet erasure probabilities, which are dependent on the OWC and RF channel conditions. Based on numerical results, it has been concluded that the increase of the number of implemented relays will not always be beneficial for system performance improvement. It is noted that the achieved  throughput gain due to relays adding  is in correlation with the channel load conditions, as well as highly  dependent on the OWC and RF channel states.

\section*{Acknowledgment}

This work has received funding from the European Union Horizon 2020 research and innovation programme under the WIDESPREAD grant agreement No 856697.

\end{document}